\documentclass{article}
\usepackage[english]{babel}
\usepackage[a4paper,top=2cm,bottom=2cm,left=3cm,right=3cm,marginparwidth=1.75cm]{geometry}
\usepackage{amsmath}
\usepackage{graphicx}
\usepackage{authblk}
\usepackage{lineno}
\usepackage{cite}
\usepackage[colorlinks=true, allcolors=blue]{hyperref}

\title{Generative adversarial network for super-resolution imaging through a fiber}
\author[1,*]{Wei Li}
\author[1]{Ksenia Abrashitova}
\author[1]{Gerwin Osnabrugge}
\author[1,2]{Lyubov V. Amitonova}
\affil[1]{Advanced Research Center for Nanolithography (ARCNL), Science Park 106, 1098 XG Amsterdam, The
Netherlands}
\affil[2]{LaserLaB, Department of Physics and Astronomy, Vrije Universiteit Amsterdam, De Boelelaan 1081, 1081 HV Amsterdam, The Netherlands}
\affil[*]{W.L.: w.li@arcnl.nl}

\date{}

\begin{document}
\maketitle

\begin{abstract}
A multimode fiber represents the ultimate limit in miniaturization of imaging endoscopes. Here we propose a fiber imaging approach employing compressive sensing with a data-driven machine learning framework. We implement a generative adversarial network for image reconstruction without relying on a sample sparsity constraint. The proposed method outperforms the conventional compressive imaging algorithms in terms of image quality and noise robustness. We experimentally demonstrate speckle-based imaging below the diffraction limit at a sub-Nyquist speed through a multimode fiber.

\end{abstract}

\section{Introduction}
Optical fibers are broadly used in many imaging applications. Multimodes fibers (MMFs) together with advanced wavefront shaping \cite{vellekoop2015feedback} enable new ways to transmit information through a hair-thin probe \cite{vcivzmar2011shaping,di2011hologram}. Nowadays, the most popular approaches for imaging through a MMF exploit transmission matrix measurements and holographic light shaping \cite{vellekoop2010exploiting,vcivzmar2012exploiting,ploschner2015seeing}. On the other hand, computational methods can provide an elegant solution for imaging without the need for active control over the light propagation. 
Compressive imaging through a MMF improves both spatial resolution and imaging speed \cite{amitonova2018compressive,amitonova2020endo,pascucci2019compressive}.
However, practical application of compressive imaging has been restricted by the strong assumption of sample sparsity and the demand to adjust the approach to different experimental conditions \cite{calisesi2021compressed}.

Recent years have witnessed the rise of deep learning as a powerful tool for computational imaging \cite{barbastathis2019use}. In particular, the generative adversarial network (GAN) has made a series of breakthroughs.
The GAN is composed of two neural networks that compete with each other during the training process. 
A properly trained GAN can generate new data belonging to a certain class of objects.
Bora et al. proposed to solve compressive sensing problem with a GAN and demonstrated sub-Nyquist imaging without any constraints on the sample sparsity  \cite{bora2017compressed}. 
In this approach, the image is produced by iterative minimization between the GAN-generated images and the measured data.
The GAN can be used to produce a clear reconstruction from the noisy $l_2$-norm solution \cite{karim2021spi,ni2021color}. Over the past years, this GAN approach for compressive imaging has seen a rapid development \cite{wu2019deep, kabkab2018task, gao2021single}. 

Recently, deep neural networks have been successfully implemented to classify and reconstruct images distorted by propagating through an MMF \cite{ PhysRevX.11.021060, Rahmani2018, chen2020binary, borhani2018learning, fan2019deep}.
However, in contrast to GAN -- which is an unsupervised machine learning framework and does not require the labeled data -- these machine learning approaches for MMF imaging require to train a network with pairs of input images and their corresponding speckle patterns at the fiber output. As a result, it complicates the pre-calibration system and requires to repeat the training procedure if the fiber configuration changes \cite{kakkava2019imaging}. Moreover, the application area has currently been restricted to coherent image transmission with an image resolution that is limited by the diffraction of light.

Here we address the problem of speckled based sub-diffraction compressive imaging through a MMF.
We propose the GAN-based framework for imaging through a MMF which does not rely on the sparsity constraint and can be easily generalized on any speckle-based measurement system. To the best of our knowledge, this work is the first experimental application of machine learning in speckle-based compressive imaging. We show that proper choice of architecture and loss functions allows for recovering images below the diffraction limit with sub-Nyquist imaging speed via an MMF. We experimentally demonstrate the superiority of our approach over standard $l_1$-norm optimization algorithms in terms of spatial resolution, noise tolerance and sensitivity to the number of measurements.

\section{Results}
\subsection{Speckle-based computational imaging}
\begin{figure}[t]
	\centering
	\includegraphics[width=0.95\columnwidth]{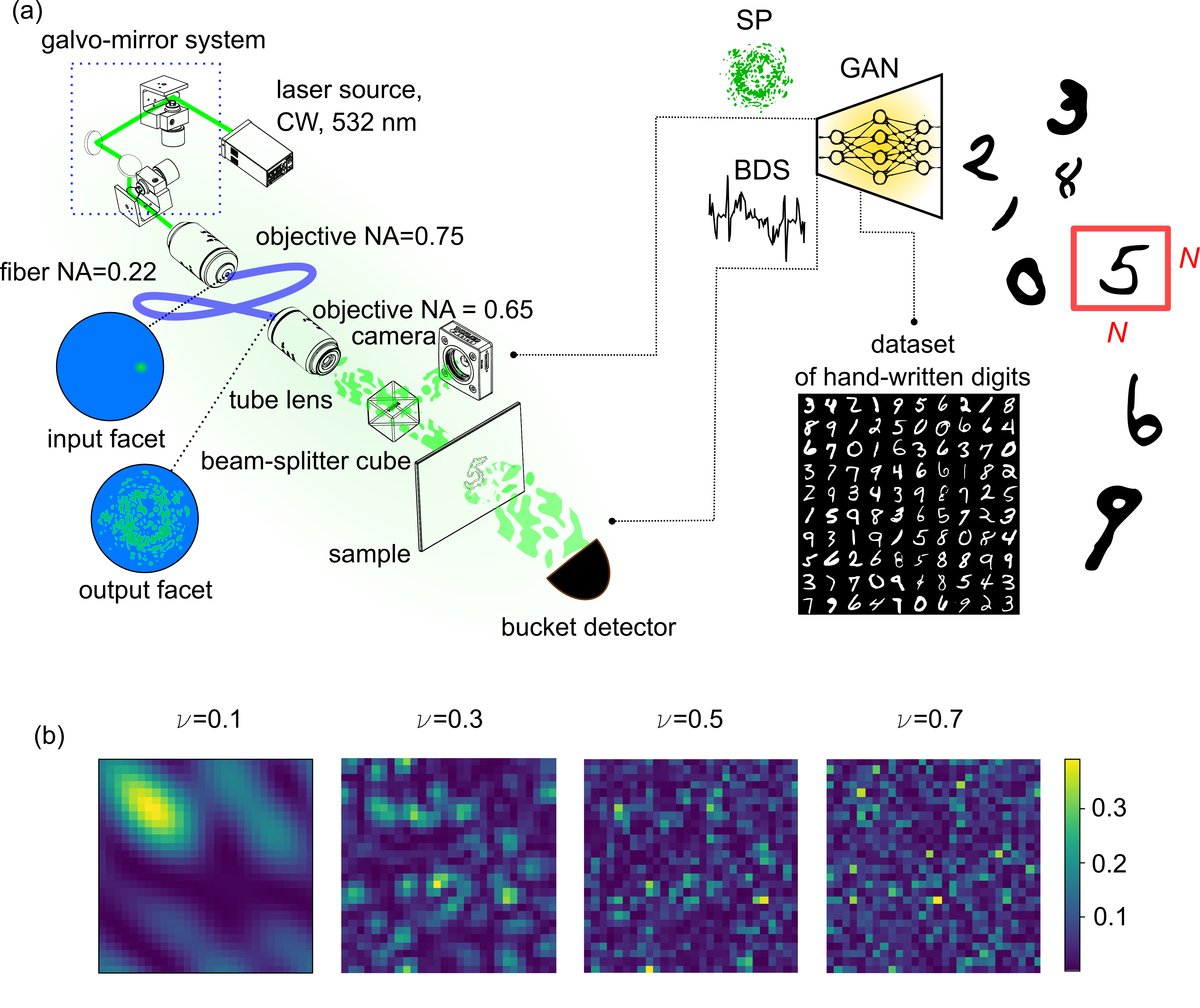}
	\caption{(a) Schematic experimental setup and representation of GAN enhanced speckle imaging. SP - speckle patterns, BDS - bucket detector signal. (b) Examples of simulated speckle patterns with different cutoff values $\nu$.
	}
	\label{fig:setup}
\end{figure}

We focus on super-resolution microscopy with a speckle illumination and single-pixel detection scheme also known as super-resolution ghost imaging \cite{amitonova2020endo}. The schematic imaging process is shown in Fig. \ref{fig:setup}(a). The sample is illuminated by a series of speckle patterns generated in a MMF or a scattering medium. 
The transmitted signal is recorded by a bucket (single-pixel) detector.
The relationship between the intensity distribution of the speckle pattern and the bucket detector signal can be represented as a linear equation, and multiple measurements can be formulated as an under-determined system of linear equations:
\begin{equation}
	\boldsymbol{y} = A  \boldsymbol{x},
	\label{CS_problem}
\end{equation}
where each flattened speckle pattern ($n\times 1$) corresponds to one row of measurement matrix $A$  ($m\times n$). The response from $m$ different speckle patterns is measured by a bucked detector to form the measurement vector $\boldsymbol{y}$ ($m\times 1$). Finally, $\boldsymbol{x}$ is the 1D unknown vector ($n\times 1$), where $n=N^2$ is the total number of pixels and $N\times N$ is the size of the 2D digitized speckle pattern. In this work, we consider the case where $m\ll n$.  

We reconstruct vector $\boldsymbol{x}$ by two computational approaches: compressive sensing algorithm and pre-trained GAN neural network.
Reconstructed vector $\boldsymbol{x}$ ($n\times 1$) is resized to represent a 2D sample image ($N\times N$).
We compare the results with the diffraction-limited images to evaluate the super-resolution and sub-Nyquist capabilities of computational approaches.
The reconstruction quality is quantified by the Pearson correlation coefficient $r$ between the reconstructed image and the truth image. The correlation $r$ ranges from $-1$ to $1$, while $1$ represents the perfect reconstruction, $-1$ represents the perfect negative reconstruction and 0 indicates that the reconstruction and ground truth are independent.

Our MMF-based single-pixel imaging setup is presented in Fig. \ref{fig:setup}(a). A detailed description of the setup and imaging methods is given in Methods section.

\subsection{Compressive imaging with GAN}
\begin{figure}
    \centering
    \includegraphics[width=0.95\columnwidth]{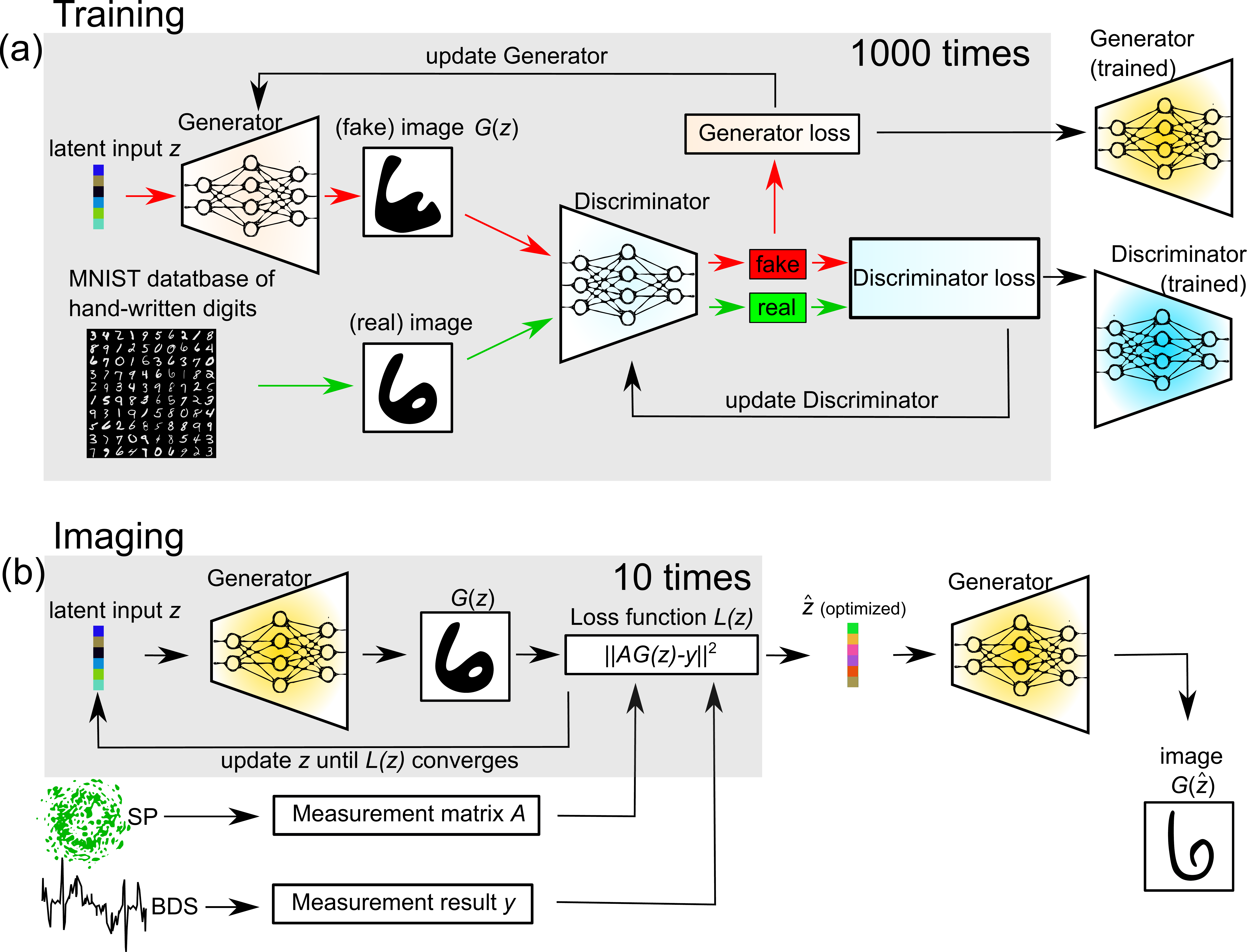}
    \caption{Illustration of speckle-based compressive imaging via GAN: (a) the training process and (b) the imaging step. SP: speckle patterns. BDS: bucket detector signal.}
    \label{fig:gan_training}
\end{figure}

The process of reconstruction by GAN consists of two main steps, as presented in Fig.~\ref{fig:gan_training}:
\begin{enumerate}
\item Training step: training the network to generate the sample images corresponding to a certain class of objects,
\item Imaging step: searching for the image that fits the experimentally measured data.
\end{enumerate}
The training step is completely independent of the reconstruction step and a properly trained network can be used for different experimental systems without any change. See Methods section for the detail.

A GAN consists of a generator and discriminator, which are simultaneously trained.
The generator network performs the transformation from a low-dimensional input latent representation $\boldsymbol{z}$ to the high resolution image, whereas the discriminator network discriminates between the images that are `real' (from the training set) and `fake' (newly generated).
These two networks learn from and compete with each other during the training process and improve their respective performance. 
When the GAN network is well-trained, the generator can generate new images belonging to the same class as the training set from an arbitrary input $\boldsymbol{z}$.

For the image reconstruction, we define a loss function $L(\boldsymbol{z})$, which can be calculated from the flattened image $G(\boldsymbol{z})$ combined with measured signal $\boldsymbol{y}$ and measurement matrix $A$:
\begin{equation}
    L(\boldsymbol{z}) = ||A  G(\boldsymbol{z}) - \boldsymbol{y}||^2.
\end{equation}

Following Bora et al. \cite{bora2017compressed}, we minimize the loss function which quantifies the discrepancy between the generator image and the experimental data. We randomly generate starting value of the input latent representation $\boldsymbol{z}=\hat{\boldsymbol{z}}$ and search for the minimum point by gradient descent:
\begin{equation}\label{eq:gradient}
    \hat{\boldsymbol{z}} \leftarrow \hat{\boldsymbol{z}} - \alpha \frac{\partial L(\boldsymbol{z})}{\partial \boldsymbol{z}}\bigg\rvert_{\boldsymbol{z}=\hat{\boldsymbol{z}}},
\end{equation}
where $\alpha$ is the learning rate. Despite the fact that the loss function is highly non-convex, we found that hundreds of iterations combined with several random initiations of the input \cite{bora2017compressed,pmlr-v80-bojanowski18a} gives a good final reconstructed signal.

\subsection{Simulations}
\begin{figure}[t]
	\centering
	\includegraphics[width=0.95\columnwidth]{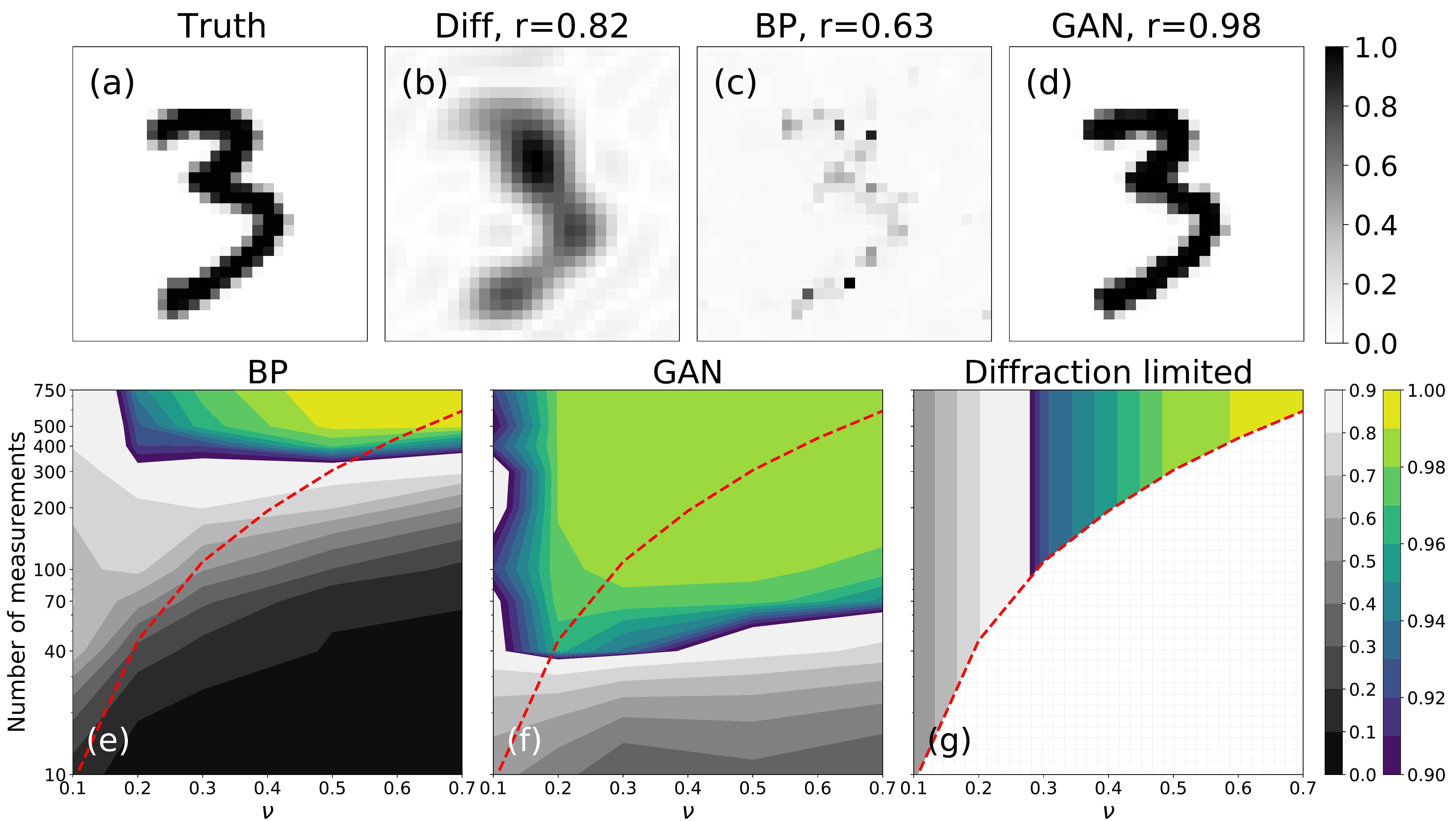}
	\caption{(a-d) An example case of the simulation results reconstructing a hand-written digit ``3'' with $\nu=0.2$ and $m\;$=$\;$100: (a) the ground truth, (b) the diffraction-limited image, (c) the BP algorithm results, and (d) the GAN results. (e) Correlation between the reconstructed and ground truth images for the BP algorithm, (f) the GAN framework and (g) the diffraction-limited case as functions of normalized spatial frequency $\nu$ and the number of measurements, averaged over the randomly selected digit of each kind ($0-9$). The gray scale bar shows the correlation range between 0 and 0.9, while the colorbar shows the detailed region from 0.9 to 1. The Nyquist limit is indicated by the red dashed line.}
	\label{fig:gan_bp_diff_measured}
\end{figure}

We first perform simulations to study the performance of the GAN imaging framework in the ideal cases without noise. 
We use the Basis Persuit (BP) algorithm for comparison and introduce the normalized spatial frequency $\nu$, which is related to the diffraction limit of the optical system. Fig. \ref{fig:setup}(b) shows examples of generated speckle patterns with different $\nu$ values. Both the BP algorithm and normalized spatial frequency $\nu$ are explained in detail in Methods section.
In Fig.~\ref{fig:gan_bp_diff_measured}(a-d), the results of the noiseless simulations performed on a hand-written digit ``3'' (randomly selected from the testing dataset) with $m=100$ and $\nu=0.2$  are presented. In these results, we compared the ground truth, the diffraction-limited results and images reconstructed with the BP algorithm and the GAN. The correlation coefficients between the reconstructed images and the ground truth are 0.98 for the GAN, 0.63 for the BP and 0.82 for the diffraction-limited approach.

We studied the performance of the reconstruction algorithms for each of the combinations of different normalized spatial frequencies $\nu=[0.1, 0.2, 0.3, 0.5, 0.7]$ and number of measurements $m=[10, 40, 70, 100, 200, 300, 400, 500, 750]$.
Each correlation coefficient between the reconstruction result and the ground truth is averaged over the digits of each kind ($0-9$) randomly selected from the testing dataset. The results are shown in Fig.~\ref{fig:gan_bp_diff_measured}(e-g).
In order to clearly show the performance difference at high $r$ region, we use two color scales.
The gray scale corresponds to $r$ range between 0 and 0.9, while the blue-green color scale shows the detailed range between 0.9 and 1. The Nyquist limit, i.e. the minimum number of measurements require for conventional imaging, is shown by the red dashed line. In the case of diffraction-limited imaging, Fig.~\ref{fig:gan_bp_diff_measured}(g), the area below the Nyquist limit is not determined.

The GAN achieves a high correlation reconstruction ($r \ge 0.9$) with a much lower number of measurements and a lower normalized cutoff-frequency than BP, as shown in Fig.~\ref{fig:gan_bp_diff_measured}(e, f).
The correlation coefficient for the GAN quickly saturates to $r=0.98$ at $\nu = 0.3$ and $m = 80$, while the BP algorithm requires $\nu = 0.45$ and $m = 450$ to reach similar performance. However, the non-convex nature of the loss function $L(\boldsymbol{z})$, used to find the optimal solution in the GAN approach, makes it hard to reach the perfect reconstruction ($r=1$). For  $\nu \leq 0.45$, both the GAN and BP demonstrate super-resolution imaging, as their image quality is better than the diffraction-limited imaging case. At the same time, both GAN and BP overcome the Nyquist criteria.
For instance, at $\nu=0.7$ we can enhance the imaging speed by roughly 2 and 11 times with BP and GAN respectively.

\begin{figure}
	\centering
	\includegraphics[width=0.9\columnwidth]{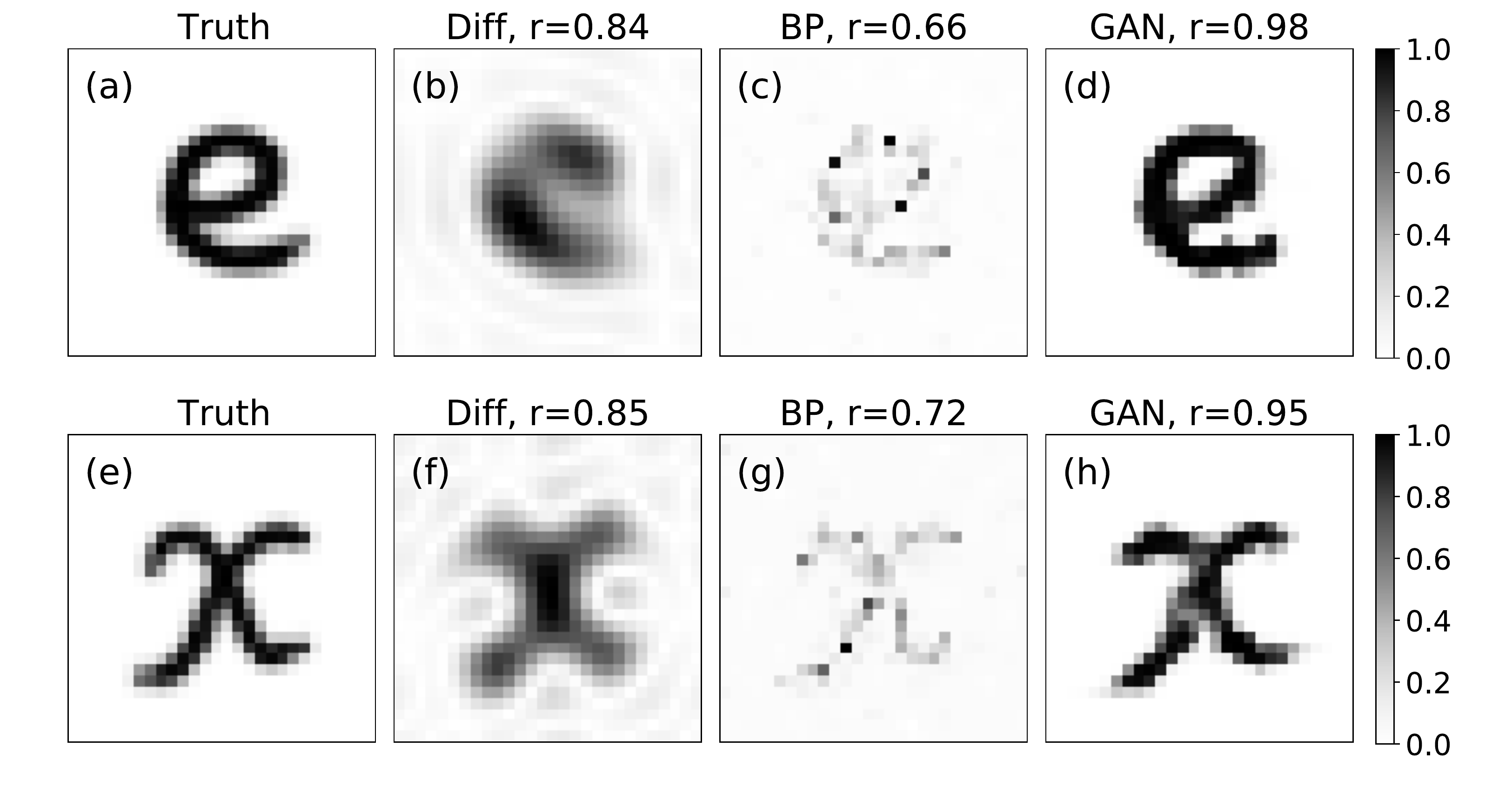}
	\caption{The results of the simulations performed on a hand-written letters ``e'' and ``$x$'' with $\nu=0.2$ and $m\;$=$\;$100: (a,e) the ground truth, (b,f) the diffraction-limited image, (c,g) image reconstructed by the BP algorithm, and (d,h) image reconstructed by the GAN.}
	\label{fig:letters}
\end{figure}

We used a different class of images to evaluate the transfer learning reconstruction performance and the flexibility of the GAN-based imaging approach. The results for handwritten letters ``e'' and ``$x$'', whose shapes are not close to the digits, are presented in Fig.~\ref{fig:letters}. Therefore, the object is not of the same class as training data, although they are both gray scale and of a handwritten nature. The imaging conditions are the same as Fig. \ref{fig:gan_bp_diff_measured}(a-d) with $\nu=0.2$ and $m\;$=$\;$100. The GAN approach in these cases show excellent performance demonstrating the flexibility of GAN-based imaging.

\begin{figure}
	\centering
	\includegraphics[width=\columnwidth]{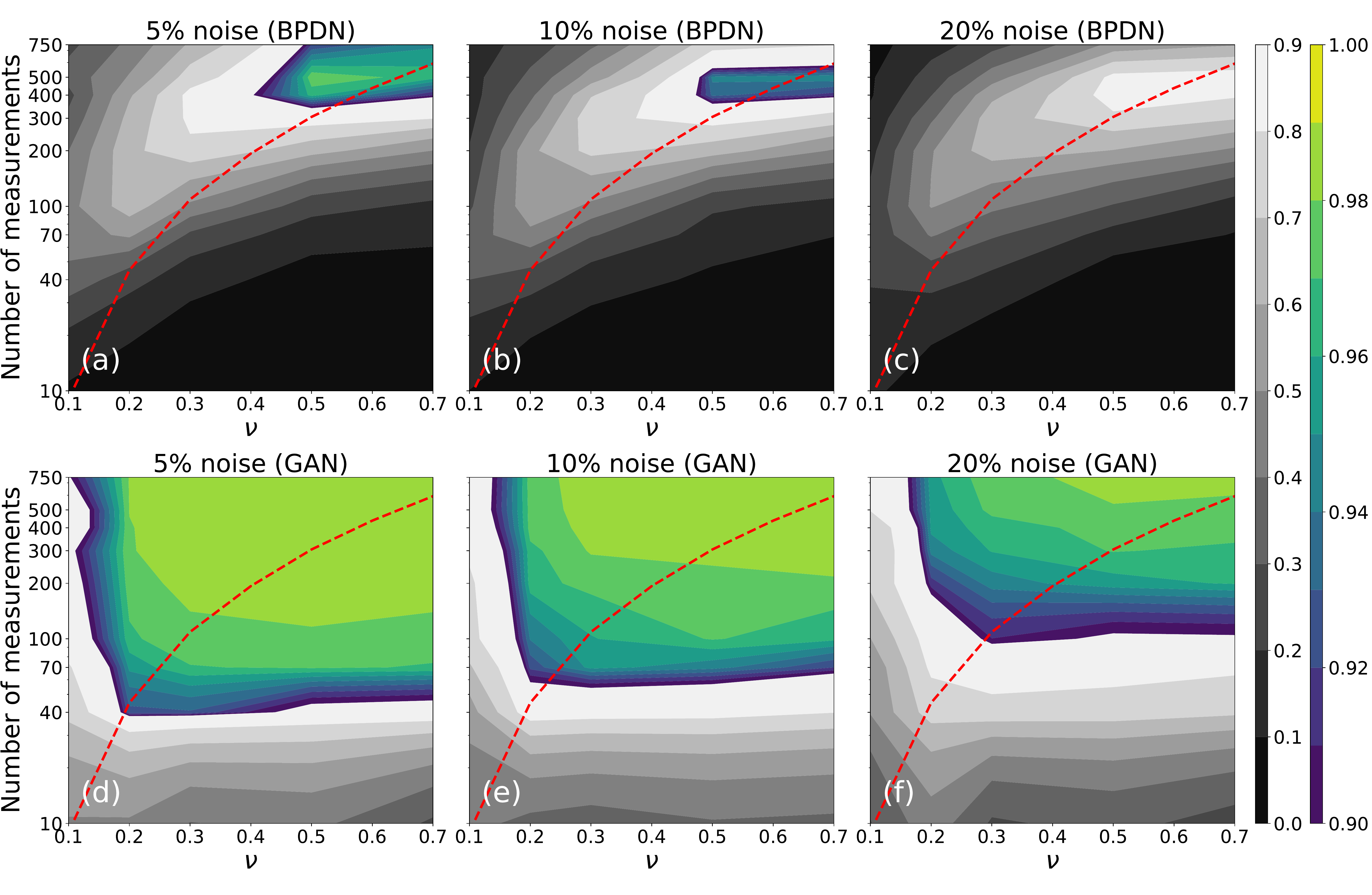}
	\caption{ Correlation between the reconstructed and ground truth images for the BPDN algorithm (a-c) and GAN framework (d-f) for different noise levels: $5\%$ (a, d), $10\%$ (b, e) and $20\%$ (c, f) as functions of normalized spatial frequency $\nu$ and the number of measurements, averaged over the randomly selected digit of each kind ($0-9$). The gray scale bar shows $0$-$0.9$ range and the colorbar shows the detailed $0.9$-$1$ region. The Nyquist limit is indicated by the red dashed line.}
	\label{fig:noise}
\end{figure}

We also investigate the performance of the proposed GAN approach under realistic conditions by adding noise on both the measurement matrix $A$ and the measured signal $\boldsymbol{y}$ as described in Section~\ref{simulation}.
For a fair comparison, we replace BP with a Basis Persuit Denoising (BPDN) algorithm, which is optimized for noisy measurements by tuning the error tolerance factor $\delta$. Meanwhile, the GAN does not require any additional parameter tuning. We repeat the reconstruction for different samples, $m$, $\nu$ and the noise levels.
The results averaged over the digits of each kind ($0-9$) are shown in Fig. \ref{fig:noise}.

The image quality of the BPDN algorithm deteriorates quickly with the increasing amount of noise, as can be seen in Fig. \ref{fig:noise}(a-c). The maximum achievable reconstruction quality of the BPDN decreases to $r_\mathrm{max}=0.8$ for $20\%$ of noise.
At the same time, the GAN provides a good reconstruction (with $r_\mathrm{max}$ up to $0.98$) even for the high noise level. The GAN approach maintains super-resolution and sub-Nyquist regimes.

\subsection{Experimental validation}
\begin{figure}[t]
	\centering
	\includegraphics[width=0.9\columnwidth]{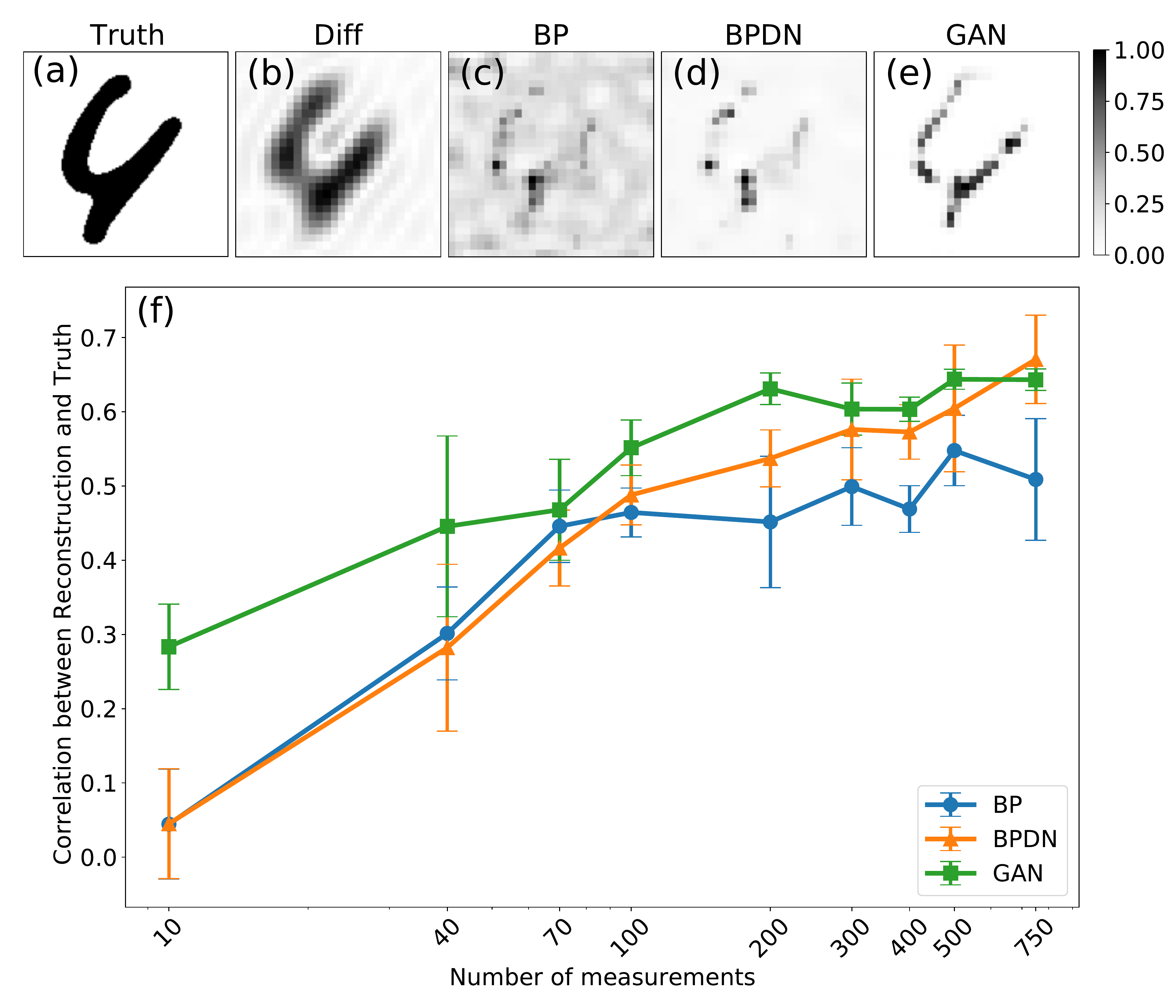}
	\caption{An example of one validation image with $m=200$. (a) The bright-field (binarized) image of the sample engraved in the aluminum film which is randomly chosen from the testing dataset. (b) The diffraction-limited image with $\nu=0.2$. (c), (d) and (e) show the reconstructed images obtained with BP, BPDN and GAN respectively. (f) The average performance of reconstruction on the real samples of BP, BPDN and GAN algorithms.}
	\label{fig:example_validation}
\end{figure}

In order to test our GAN reconstruction framework in practice, we performed an MMF-based compressive imaging experiment on a real sample. Handwritten digits from the testing subset of MNIST database are prepared on a microscope slide using mask-less UV-photolithography (365~nm) and lift-off of sputtered reflective 200~nm thick aluminium film. To generate the ground truth, the separately measured bright-field image of the handwritten digit ``4'' (randomly selected from the testing dataset) engraved in the aluminum film is cropped and converted to the binary format, as presented in Fig.~\ref{fig:example_validation}(a).

The detailed description of the experimental setup is provided in Section~\ref{setup}. Given the diffraction limit of our experimental setup, the effective normalized spatial frequency corresponds to $\nu\approx0.2$. The diffraction-limited image, calculated from the ground truth, obtained by applying a low-pass filter with $\nu=0.2$, is presented in Fig.~\ref{fig:example_validation}(b).
We perform $m=[10, 40, 70, 100, 200, 300, 400, 750]$ random measurements by scanning the input facet of the MMF and randomly choosing $m$ speckle patterns and $m$ corresponding detected intensities to process.
The recorded speckle patterns are cropped and resized to match the size of the dataset images.
The processed speckle images form measurement matrix $A$.

Figure \ref{fig:example_validation}(c-e) shows the imaging results for $m=200$ random measurements and the reconstruction performed by the BP algorithm (c), the BPDN algorithm (d), and the proposed GAN (e).
The GAN framework provides significantly better image quality compared to diffraction-limited microscopy and traditional compressive sensing algorithms. The GAN approach has much more resemblance to the ground truth and contains a noiseless background.

To characterize the imaging performance, the correlation coefficients between the reconstructed images and the ground truth have been calculated and are presented in Fig.~\ref{fig:example_validation}(f) for the different number of measurements $m$. For each $m$, we repeat the experiment and the reconstruction procedure five times and average the correlation coefficient over all realizations, calculating the mean and the standard deviation. For the BPDN the error tolerance factor $\delta$ is tuned to achieve the best quality of reconstruction for each number of measurements.
As expected, for all the algorithms the quality of reconstruction increases with the number of measurements.
The GAN approach shows better performance compared to the BP and BPDN algorithms for $m<750$, which is in full agreement with the results of the simulation.
The BPDN algorithm performs better than BP, which can be explained by the fact that real measurements always contain noise that is not taken into account in the BP formulation Eq. \ref{BP}. 
The BPDN demonstrates a slightly better quality of reconstruction compared to the GAN for $m=750$, because of the non-convex nature of the GAN loss function.

\section{Discussion and Conclusion}
In this study, we experimentally demonstrated compressive fiber imaging with a deep convolutional GAN framework. State-of-the-art methods of machine learning allow to reconstruct images distorted by the MMF only in the case of coherent imaging \cite{Rahmani2018, chen2020binary, borhani2018learning, fan2019deep, kakkava2019imaging}. Whereas the proposed GAN approach has no restrictions and can be used for popular incoherent methods of fiber imaging such as (auto-)fluorescent microscopy. Moreover, the GAN framework does not require to repeat a training procedure with pairs of input and output images for different fiber configurations.

With the proposed GAN approach, we demonstrated fiber-based compressive imaging with enhanced spatial resolution. We have shown that the imaging quality outperforms diffraction-limited imaging approaches. In contrast to conventional minimization algorithms typically used for reconstruction of compressed data with sub-diffraction resolution, our approach does not require the samples to be sparse. The sparsity constraint is very general and can, in principle, be applied to many natural images. However, finding the sparsity domain for a certain sample is not always straightforward. In addition, the imaging quality significantly deteriorates with the increased amount of noise and for lower levels of sample sparsity \cite{Lochocki2021}. 

We theoretically and experimentally demonstrated that the proposed GAN-based computational framework can achieve an enhanced quality of reconstruction compared to the traditional compressed sensing minimization algorithms. These results are more pronounced for the small number of measurements, while for the large number of measurements the GAN and the traditional BPDN perform similarly. The GAN is also proved to be tolerant to a large amount of noise.
Moreover, the GAN in this study is capable of reconstructing images of handwritten letters as well. Hence the well trained GAN can also be applied to reconstruct images from other testing datasets, provided the samples have similar features.

The GAN requires a properly tuned network architecture and the training process of the GAN demands a lot of computational power. The computational complexity can be reduced by either reducing the size of the training set or increasing the size of the latent space.
However, both procedures may deteriorate the performance of the generator. In our study the size of the latent space is 100, which is less than an order of magnitude smaller than the overall number of pixels $n=784$. We have also done the simulations with the latent space size of 50 and it also gives similar promising results.

No additional training is needed to make the transition from one experimental setup to another, as the GAN is trained to images without experimental imperfections. The approach can be easily generalized to any speckle-based single pixel detection setup of imaging without re-training the GAN. As a result, the proposed GAN approach paves the way towards broad implementation area with a sub-diffraction imaging quality not limited by a sparsity of a sample. 

\section{Methods}
\subsection{Experimental setup}\label{setup}
The experimental setup is illustrated in Fig.\ref{fig:setup}(a). The light from a laser source (532$\,$nm, Cobolt Samba, continuous wave) passes through a half-waveplate and a polarizing beamsplitter cube to control the incident power.
The laser beam is reflected by a pair of galvo-mirrors that are
projected on the entrance aperture of the objective (NA$\;$=$\;$0.75, Olympus) by relay lenses.
The objective focuses the beam on the input facet of the MMF (NA$\;$=$\;$0.22, diameter, $d=\;$50 $\mu$m). The number of modes in the MMF for single polarisation is $M_\mathrm{modes}=1050$.
Galvo-mirrors move the focus along the input facet, which excites different sets of propagating modes and produces different output speckle patterns $I_{i}$. 
The $54\times$-magnified image of the speckle pattern is projected on the sample and the camera (pixel size$\;$=$\;$2.4 $\mu$m, Basler acA 3088-57 um) by the objective (NA$\;$=$\;$0.65, Olympus), a tube lens ($f =\;$200 mm) and a beamsplitter.
The total intensity transmitted through the sample is measured by the avalanche photodiode (Thorlabs APD410A2) and forms measurement vector $\boldsymbol{y}$. The camera and the galvo mirrors are triggered by a data acquisition board (NI-PCI 6353) and controlled by a custom software.

\subsection{Simulation of speckle-based imaging}\label{simulation}
As a first step, we perform simulations to evaluate the performance of the reconstruction algorithms for different number of measurements, spatial frequencies and levels of noise. We generate a random measurement matrix $A$ by creating $m$ random well-developed speckle fields (of size $n = 28\times28$ pixels) which are sampled from a circular Gaussian distribution \cite{goodman2015statistical}. The intensity distributions of these fields form the rows of our simulated measurement matrix. 
To study the diffraction limit effects, we apply a low-pass filter with a cutoff frequency $\nu$ to the random field  distribution, where $\nu$ is normalized to the maximum spatial frequency in the image. As $\nu$ decreases the speckles get larger, hence less spatial information is obtained from the measurement. 

As a sample, we use handwritten digits from the standard \textit{MNIST} database \cite{lecun1998gradient}, which contains $70000$ images ($60000$ in the training dataset and $10000$ in the testing dataset) with an image size of $28\times 28$, hence $n=784$. The average sparsity of the complete dataset is 0.19.

In the noiseless case, the measurement signal from the simulated bucket detector is simply given by $\boldsymbol{y}=A\boldsymbol{x}$, where $\boldsymbol{x}$ is the flattened ground truth image of a handwritten digit. Different levels of noise (5\%, 10\% and 20\%) are added to both the measurement matrix and the measured signal. The noise follows a Gaussian distribution with a mean of zero and the standard deviation that equals the standard deviation of the corresponding measurement matrix or measured signal.

\subsection{GAN training}
For our deep convolutional Generative Adversarial Network \cite{radford2015unsupervised}, we modified the design and the code from \textit{TensorFlow} \cite{dcgan2021}. The \textit{Keras Sequential API} is used to define both the generator and the discriminator architectures. 
The input latent representation $\boldsymbol{z}$ is a ($100\times1$) vector whose entities follow the standard normal distribution. To transform $\boldsymbol{z}$ to the image ($28\times 28$), the generator uses the transposed convolution (deconvolution) layer several times, with the \textit{LeakyReLU} activation for each layer (output layer uses hyperbolic tangent activation function (\textit{tanh})).
The discriminator is a convolutional neural network based image classifier, which gives positive values for real images and negative values for fake images.

Fig. \ref{fig:gan_training}(a) shows the detailed training process of the deep convolutional GAN. The training dataset, which contains 60,000 handwritten digits, is shuffled and separated into smaller batches with the size of 256 to reduce computational complexity. The generator generates 256 fake images based on $\boldsymbol{z}$. 
Discriminator maps these 256 fake images and 256 real images to either positive (real output) or negative values (fake output). 
Fake output is used to calculate the loss function for the generator, while both the fake output and the real output are used to calculate the loss function of the discriminator. 
The generator and discriminator are updated by gradient descent with the calculated loss functions, implemented by the \textit{Adam} optimizer \cite{kingma2014adam} with the learning rate of 0.0001. The process of training for the whole training set is repeated 1000 times.

\subsection{Reconstruction with GAN algorithm}
Fig. \ref{fig:gan_training}(b) shows the detailed reconstruction algorithm. The random input vector $\boldsymbol{z}$ ($100\times1$) goes through the generator, which generates the corresponding image $G(\boldsymbol{z})$ ($28\times28$). 
In order to implement the gradient descent method, we use \textit{GradientTape} to do the automatic differentiation and use \textit{Adam} optimizer \cite{kingma2014adam} with the learning rate of 0.1 to update the input latent representation $\boldsymbol{z}$. We use 2000 iteration steps in the gradient descent optimization and repeat the procedure of optimization 10 times with new randomly generated starting point $\boldsymbol{z}$.
We select the final reconstruction result $G(\hat{\boldsymbol{z}})$ out of 10 by choosing the $\hat{\boldsymbol{z}}$ with the lowest values of loss function $L(\hat{\boldsymbol{z}})$.

\subsection{Basis Pursuit}
Basis pursuit algorithm (BP) can solve the compressive sensing problem by
\begin{equation}
	\min_{\boldsymbol{x}}||\boldsymbol{x}||_{1} \quad \textrm{s.t.} \quad \boldsymbol{y} = A\boldsymbol{x},
	\label{BP}
\end{equation} 
where $||\boldsymbol{x}||_{1}=\sum_{i}|\boldsymbol{x}_{i}|$ is the $l_{1}$ norm of vector $\boldsymbol{x}$ or the sum of the absolute values of vector entities.
In the case where noise is present, the BP algorithm is substituted by the BP denoising (BPDN) algorithm: 
\begin{equation}
	\min_{\boldsymbol{x}}||\boldsymbol{x}||_{1} \quad \textrm{s.t.} \quad ||\boldsymbol{y} - A\boldsymbol{x}||_2^2 \leq \delta,
		\label{BPDN}
\end{equation}
where $\delta$ is the error tolerance factor which is dependent on the amount and nature of noise in the measurement system, the number of measurements $m$ and the number of non-zero elements (sparsity) of vector $x$. The $||.||_2$ is the euclidean norm. The BP denoising becomes BP when $\delta$=0.
We implement the BP algorithm in this study with the \textit{spgl1} package \cite{van2011sparse,spgl12021}.

\section*{Acknowledgements}
Part of this work has been carried out within ARCNL, a public-private partnership between UvA, VU, NWO and ASML and was partly financed by `Toeslag voor Topconsortia voor Kennis en Innovatie (TKI)' from the Dutch Ministry of Economic Affairs and Climate Policy. We acknowledge the support from Nederlandse Organisatie voor Wetenschappelijk Onderzoek (WISE). We thank Sergey Amitonov (TU Delft) for custom-made samples, Mark Mol (ARCNL) for his support in constructing the setup and Marco Seynen (AMOLF) for his help in programming the software.

\bibliographystyle{unsrt}
\bibliography{mybib}

\end{document}